\documentclass[12pt,epsf]{article}
\usepackage{epsfig,graphics,color}
\author{\large \bf  Z.~Usubov\footnote                                    
         {On leave of absence from Institute of Physics, Baku, Azerbaijan}
\\
\\Joint Institute for Nuclear Research,
\\ Dubna, Russia}          

\title { Looking for Squark Pair Production in the \\      
                           Early LHC Data }              

\begin{document}
\maketitle
{
\vskip 0.5cm
\hskip 6.5cm {\bf \large { Abstract}}
\vskip 1.0cm
$ $

We examine the ability of LHC experiments to observe jets
from squark pair production at the center-of-mass energy
 E$_{CM}$=10 TeV and
1 fb$^{-1}$ of integrated luminosity. 
We point out 
the crucial influence of initial- and final-state 
radiation on the signal/background discriminating ability
of different kinematic variables.
The reliable measurements of 
missing transverse energy 
and stransverse mass would play a key role in picking out the
signal against  the background.
\vskip 0.5 cm


\large {
\section{Introduction}
$ $

Among many exciting challenges that   the                          
Large Hadron Collider (LHC) era  brings 
most difficult  is to build of the fundamental theory 
to describe the physics at arbitrary high energies.
Supersymmetry (SUSY)\cite{susy1} is regarded as 
a widely favored candidate.

SUSY with  beautiful mathematical grace           
is good almost in all senses:    
the quadratically divergent contributions to the scalar sector are
canceled, the light Higgs boson is predicted,
the SUSY particle spectra contain the dark matter candidates.              
SUSY allows the unification of all known fundamental forces and,
presumable,  allows explanation of baryon asymmetry in the universe,
superstring theory naturally incorporates SUSY.

To  exclude some unwanted superpotential terms
and  provide  the proton longevity
the SUSY theories are usually supplied  with discrete symmetry --              
R-parity conservation\cite{faye}\footnote
{The KK-parity conserving  universal extra
dimension\cite{ued1} models  and  little Higgs\cite{lith} models with 
conserved T-parity 
as  extensions of the Standard Model, predicts their  own
candidates for dark matter.}.                         
As a consequence, 
the SUSY  superpartners of the Standard Model (SM) particles are
produced in pairs and decay to a SM particle and 
the lightest supersymmetric
particle (LSP). The LSP is neutral, stable and weakly interacting 
and usually  identified as the lightest neutralino  or gravitino
escaping the detector unseen.

Unfortunately,
the origin of SUSY breaking is not enough explored  yet. 
The parametrization of  soft SUSY breaking in supersymmetric
Lagrangian density  is now the basis of
many models used in SUSY searches in 
collider experiments. Among them are the  models with
graviton, gauge and anomaly mediated  soft SUSY breaking terms.
Soft parameters can be determined from the  LHC and ILC data if
SUSY is discovered.

The most popular SUSY breaking mechanism is realized in
minimal supergravity (mSUGRA)\cite{msug1}, where the SUSY  
breaking mediated from "hidden" to the "visible" sector by
gravitational interactions.
These    models are strongly motivated, consistent with
experimental data, and simple enough                     
to explore in current collider experiments\footnote {      
About some problems in mSUGRA 
and its solutions see e.g. Ref.~\cite{bae2}.}.          
In mSUGRA  five 
parameters, four continuous and one discrete,  
$$ m_{0}, m_{1/2}, A_{0}, tan{\beta}, sign({\mu}),$$
are enough to calculate superpartner masses and mixings.
Here $m_{0}$ is the  common mass for scalars, $m_{1/2}$ is the common
gaugino mass,
$A_0$ is the common soft trilinear SUSY breaking parameter,
$sign({\mu})$  is the sign of the Higgsino mass term, $tan{\beta}$    
is the ratio of  the vacuum expectation values of two Higgs doublets
giving mass to the up and down type quarks. The   parameters 
$m_{0},\, m_{1/2}$,  $A_{0}$ 
are defined at the grand unification scale and  $tan \beta$ at the
electroweak scale.
Note that   low-energy parametrization of the 
Minimal Supersymmetric Standard Model                         
contains up to  120 parameters arising mainly in the 
soft symmetry breaking terms.

Intensive searches for manifestation of SUSY  at the 
Tevatron   were performed by CDF and D0 collaborations
(see  e.g. Ref.~\cite{cdfd0s} and refs therein). The LEP and HERA
Collaborations also  looked for SUSY in a  variety
of channels (see Refs.~\cite{lep1,hera}).

The ATLAS and CMS experiments have developed search 
strategies\cite{atl1,cms1} covering 
different SUSY breaking schemes and event topologies.

At the hadron colliders 
the inclusive  production of squarks and gluinos
 via strong interaction
dominate over associated production of charginos and 
neutralinos and is
one of the most promising discovery channels for SUSY.

In this paper, we restrict ourselves to the study of the 
squark pair production in $pp$ collisions at the 
center-of-mass energy  $E_{CM}=10$ TeV.
This channel suffers from high backgrounds from QCD jet events.
The requirement of high missing transverse energy values 
effectively rejects this background. 
We will also focus    on 
the azimuthal angle between
two hardest jets, the variables $m_{T2}$\cite{lest} and
$\alpha$\cite{rand}.
To eliminate multistep cascade decays of squarks we choose    
the  parameter  point of mSUGRA where  squarks are lighter than
the gluinos. In this case, the primary decay modes are 
$\tilde q \to q \tilde \chi_{i}^0$. 
Thus, the squark pair will be presented in the interaction products
as two quarks and two neutralinos.
The final state exhibits missing energy and no mass peaks can
be directly reconstructed.

In addition to the QCD  events, 
the main background to SUSY dijet signatures  are
electroweak processes such as
$W+jets$, $Z+jets$, $tt+jets$ events. We find that the diboson
production rate in the background is negligibly small for
this study.

The rest of this paper  is organized as follows. The next section 
gives a brief description of signal and background               
events, the choice of the SUSY benchmark point and some parameters
of the  generic LHC detector.                                              
Section~3 gives our strategy for the search for SUSY        
dijet events. We make use of different kinematic variables and
examine signal and background events. In this Section we 
also make an estimate of the $\tilde q \bar {\tilde q}$ 
signal-to-background ratio. The impact  of the calorimeter
energy resolution,  initial- and final-state radiation 
and jet reconstruction algorithm to the 
dijet variable distributions are studied.
We end with the conclusions
in Section~4.
\newpage
\section{Signal and Standard Model  background simulations for the 
generic LHC detector}
$ $

In order to study the ability of the LHC 
to observe the SUSY dijet events at $E_{CM}=10$ TeV  we
used the PYTHIA6.4\cite{pyt1} event generator
for signal and QCD background simulation.
The ALPGEN2.12\cite{alpg} code was used for $tt+nj$, $W+nj$,    
$Z+nj$ and diboson samples, where $nj$ denotes  $n$ light jets.
It was interfaced with PYTHIA for subsequent jet 
showering and hadronization.
We include the following processes 
in our background analysis:
0-2 jets in $tt$, 0-3 jets in $Z$ 
and 0-4 jets in $W$ events.
Only in one diboson (ZZ, ZW, WW) event 
for 1 fb$^{-1}$ of data set the sum of the transverse momenta  
of two hardest jets  
becomes  greater than 500 GeV (see below).

For this analysis, we used   the mSUGRA scenario with 
the commonly used SPS1a'\cite{sps1}
benchmark point. The model parameters at this point are $m_0=70$~GeV,
$m_{1/2}=250$~GeV, $A_0=-300$~GeV, $tan \beta =10$, $\mu>0$.
We used the SPheno\cite{sphe} package for calculating
the  sparticle mass spectra, decay widths and  branching ratios.
The benchmark point used  lead to a SUSY particle spectrum with
$m_{\chi^{1}_{0}} \sim$98 GeV, 
$m_{\tilde q}  \sim$550 GeV, 
$m_{\tilde g} \sim$608 GeV.
The total leading-order 
cross section of squark pair production at this
point  is $\sim$4.65~pb 
if the transverse momentum of the outgoing partons 
in the hard scattering process is greater than 50 GeV.

We use the central values of leading-order 
parton distribution function set 
from  CTEQ6L1\cite{pump}            
and do not change the PYTHIA6.4
default choices for $Q^2$ definition as well as
factorization/renormalization scales.
The initial-  and final-state 
QCD and QED radiation (IFSR) 
and multiple interactions (MI)  were enabled.

The detector performance was simulated by using the 
publicly available PGS-4\cite{conw} 
package written by J.~Conway and modified
by S.~ Mrenna for  the generic  LHC detector.                         
The calorimeter granularity is set to 
$(\Delta \phi \times \Delta \eta)=(0.10 \times 0.10)$. Energy                  
smearing in the hadronic calorimeter of the generic 
LHC detector is governed by\footnote {We add the constant term
to the PGS-4 simulation of energy smearing
in the hadronic calorimeter.}
$$
  {\Delta E \over {E}} = {{ a  \over \sqrt{E}} \oplus b   } 
\qquad (E\,\,in\,\,GeV),
$$
where the stochastic term factor is $a=0.8$ and the 
constant factor is  $b=0.03$.
Jets were reconstructed down to $|\eta|\le 3$ using 
the  $k_T$ algorithm implemented in PGS-4. We chose $ D=0.4$
for the jet resolution parameter and required that both leading      
jets carried a transverse momentum $p_{T}^{1,2}>50\,$GeV.

In order to suppress backgrounds from the semileptonic 
SM and SUSY processes we select events 
without  any isolated muon, electron, tau or photon
with $p_T>20$ GeV. We use the simplified
output from PGS-4, namely,  a list of two most energetic
jets.

We simulated signal and background events at the rates
corresponding to 1~fb$^{-1}$ of accumulated data.                       
\begin{figure}[h!]                             
{\vskip -2.0cm}
{\hskip  0.0cm}  {\epsfxsize  6.0 truein \epsfbox{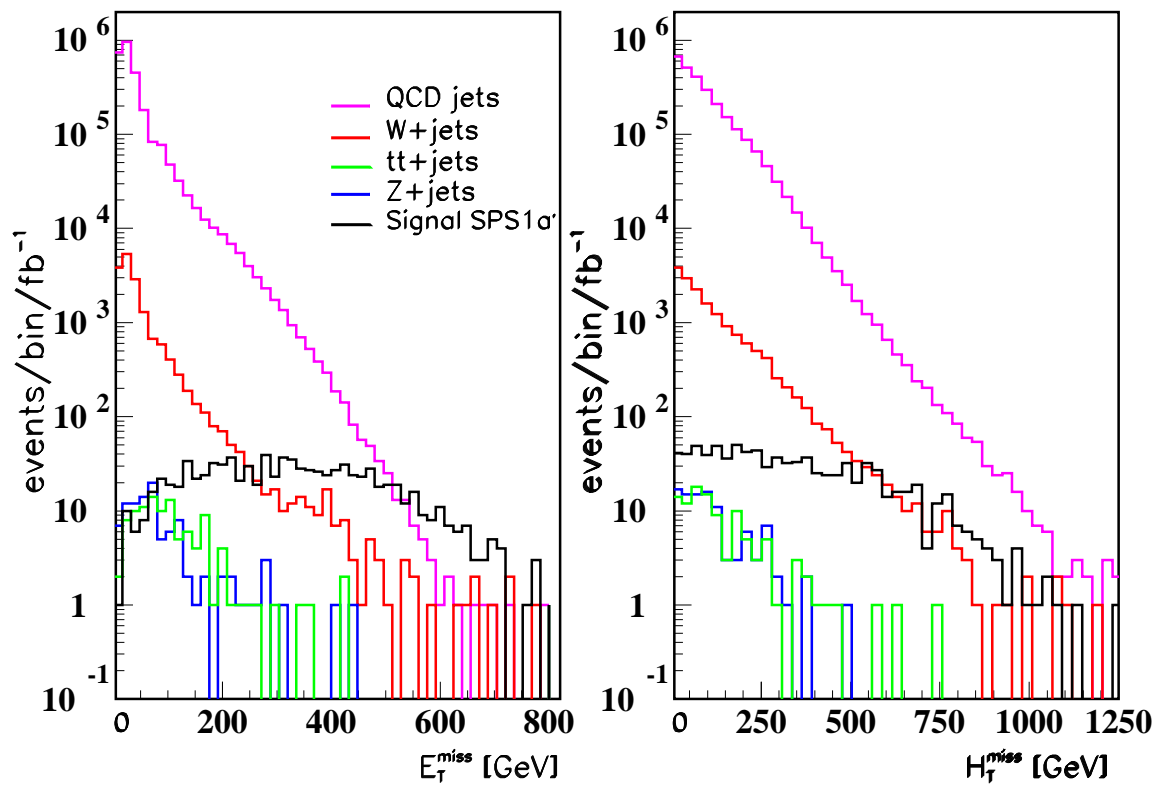}}
{\vskip -1.5cm}
\caption[]{Distributions of the $E_{T}^{miss}$  (left) and $H_{T}^{miss}$ (right) 
variables (see the text) for the signal and background 
events at $E_{CM}=10$ TeV 
and integrated luminosity of 1 fb$^{-1}$ and the SPS1a' benchmark point.
The sum of the transverse momenta of two 
hardest jets satisfies $p_{T}^1+p_{T}^2> 500$~GeV.} 
\label{Norm}
\end{figure}

We note that the K-factor 
is $\sim$1.8-2.0  for QCD background ,  
$\le$1.2             for $W+jets$ and
$Z+jets$\cite{camp}, 
$\sim$1.1     for $tt+1j$\cite{ditt},  and 
$\sim$0.89     for $tt+2j$\cite{wore}  events.
The K-factor for the signal events at the benchmark  point  
SPS1a' calculated with
using Prospino2.1\cite{prosp} is about 1.5. 
In this analysis no K-factor was applied 
and we recognize that our results 
on signal significance may be  overestimated.
\begin{figure}[h!]                            
{\vskip -3.0cm}
{\hskip  0.0cm}  {\epsfxsize  6.0 truein \epsfbox{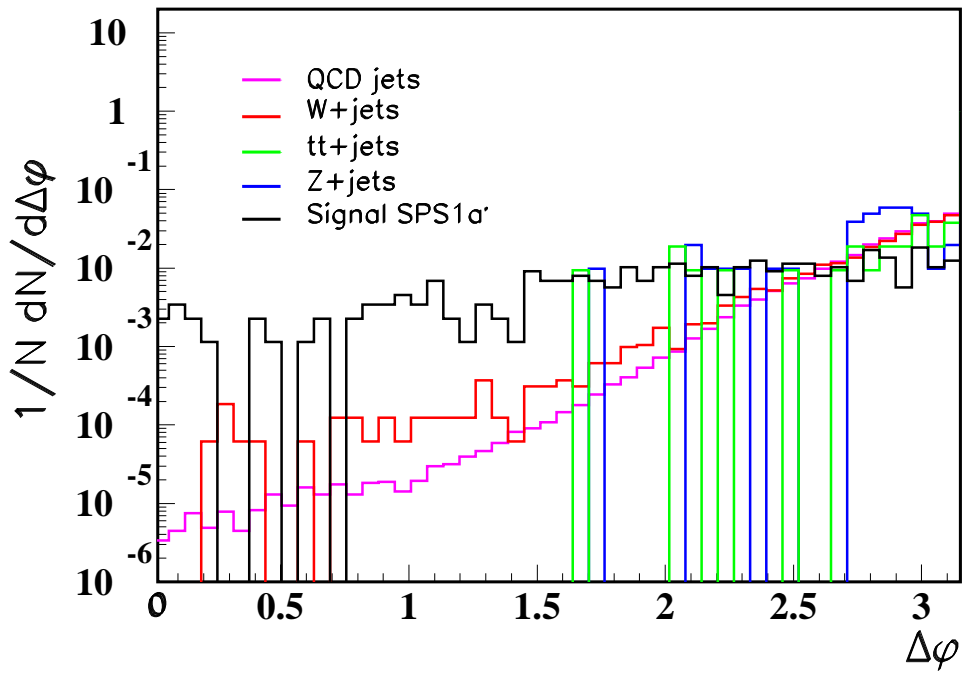}}
{\vskip -2.5cm}
\caption[]{The normalized distribution of the  
difference of the  azimutal angles
for two hardest jets ($p_{T}^1+p_{T}^2>500$ GeV) 
for signal events along with various Standard Model
background sources at $E_{CM}=10$~TeV and 1~fb$^{-1}$ of integrated
luminosity.}
\label{Norm}
\end{figure}

\begin{table}[]
\large {
\begin{center}
{\color{black}        
\begin{tabular}{||c||c||c||c||c||c||c||}       \hline \hline
$ E_{T}^{miss}\,[GeV] $  & 200  & 300  & 350 & 400 & 450 & 500  \\ \hline 
$S/B $               & 0.02 & 0.09 & 0.19& 0.45&0.94 &1.79 \\ \hline \hline 
\end {tabular}
\caption{The signal-to-background ratio for the $\tilde q \bar {\tilde q}$ 
production in  pp interactions at $E_{CM}=10$~TeV and 1~fb$^{-1}$
of integrated luminosity estimated at different selection
cuts for  $E_{T}^{miss}$.}
}
\end{center}
}
\end{table}

\begin{figure}[h!]                            
{\vskip -2.2cm}
{\hskip  0.0cm}  {\epsfxsize  6.0 truein \epsfbox{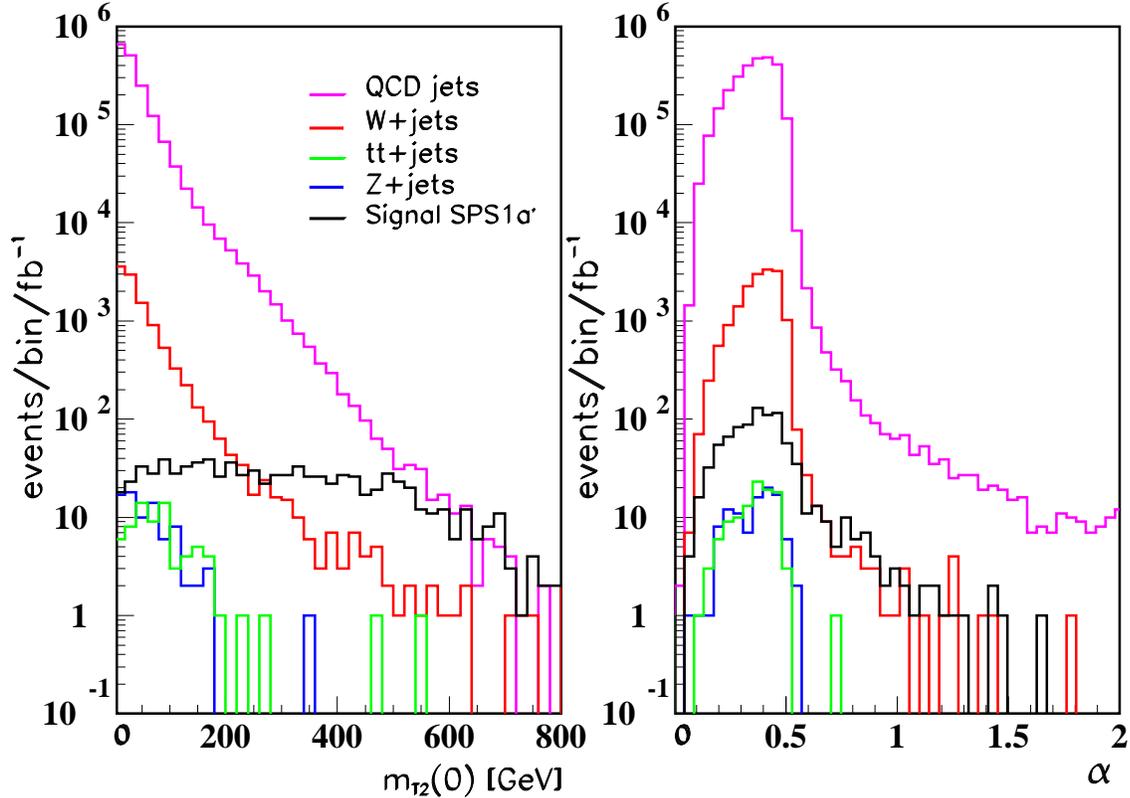}}
{\vskip -2.0cm}
\caption[]{Expected distributions of the  variables 
$m_{T2}$  (left) and $\alpha$
(right)
for the signal and background events at $E_{CM}=10$ TeV 
and integrated luminosity of 1 fb$^{-1}$ and SPS1a' benchmark point.}
\label{Norm}
\end{figure}

\begin{figure}[h!]                            
{\vskip -3.5cm}
{\hskip  0.0cm}  {\epsfxsize  6.0 truein \epsfbox{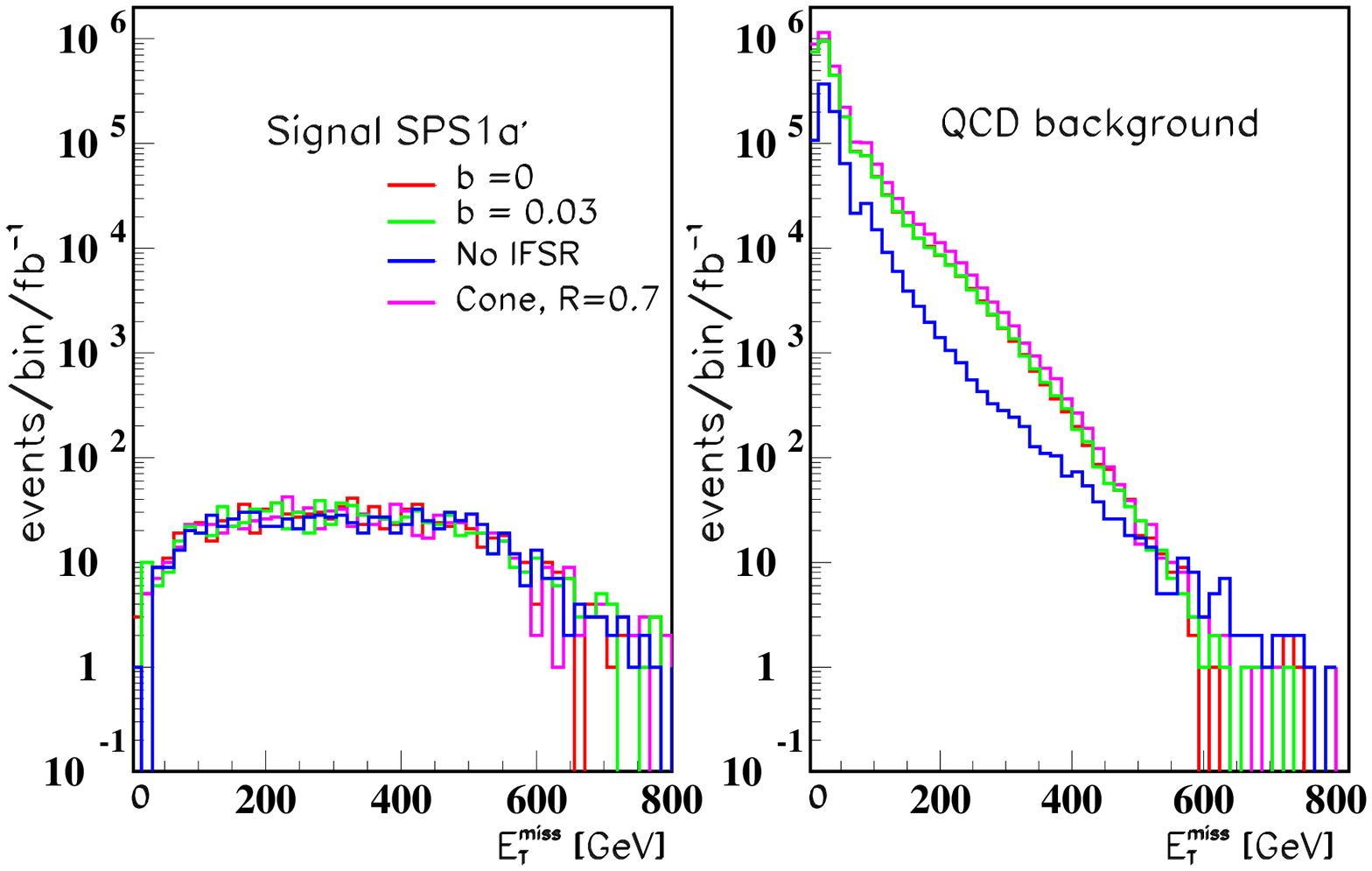}}
{\vskip -2.0cm}
\caption[]{Expected $E_{T}^{miss}$ distributions for signal 
events at the SPS1a' benchmark  point (left)
and QCD background events  with and without initial- and  
final-state radiation, for 
different constant terms $b$ in the 
hadron calorimeter resolution expression,
and with the jet cone algorithm (right).
LHC operation at $E_{CM}=10$ TeV and
integrated luminosity of 1 fb$^{-1}$ is assumed.}
\label{Norm}
\end{figure}

\section{Looking for squark pair production for one particular 
mSUGRA benchmark point}
$ $

Even in the early stages of its operation the LHC allows one
to reach very large values of the jet transverse energy,   the region
which has never been studied before.

We start with comparing the 
missing transverse energy  $E_{T}^{miss}$ of the whole event 
and  $H_{T}^{miss}$   defined using only two leading jets 
in the event.
In Fig.~1 we plot signal and background events binned in 
$E^{miss}_T$  and $H_{T}^{miss}$.
The only  selection criterion 
requires a hard cut of 500 GeV on the sum 
of the transverse momenta of two hardest jets.
A comparison of the left  and right panels of Fig.~1 shows that 
the effectiveness of background restrictions
based on  the $E_T^{miss}$ and $H_T^{miss}$ 
cuts would significantly differ towards  high values 
of these  variables.
In Table~1 we present the signal-to-background ratio, S/B, for
different selection cuts for $E_T^{miss}$.    
It would be possible
to achieve the signal significance\footnote {
We follow the tradition but
can say that if no robust signature in the mass peak is observable,
the use   of this   signal significance is not so compelling.}
$S/\sqrt{S+B}$ of about 10.5 
with the accumulated data of 1 fb$^{-1}$
using only the $E_{T}^{miss}$ cut. 
With only the  cuts on $H_{T}^{miss}$ 
it is impossible   to get signal significance even close to 3.

Besides being of physics origin, 
events with the missing transverse energy  also have 
other sources. Mismeasurements of the energy of jets,
incomplete coverage of the calorimeters, large
electronic noise, etc. would lead to a large $E_{T}^{miss}$
signal. For this reason, in the early stage of the LHC running
it may not be possible to use $E_{T}^{miss}$ as a 
signal/background discrimination variable. In what follows  
we  do not implement any
$E_{T}^{miss}$ requirement.

The dijet angular distribution is a useful measurement 
to probe the SM processes as well as new physics manifestation.
In Fig.~2 we plot the normalized dijet angular distribution,
$1/N\,dN/d\Delta \phi$, where 
$\Delta \phi$=${\phi}_{1} - {\phi}_{2}$ for two leading jets.
The signal event distribution is nearly flat in contrast to
the background event distributions.

An important quantity which in principle
allows  observing squarks and  determining     
them masses  and masses of  their invisible decay products 
is the $m_{T2}$\cite{lest}.                                           
Inspired by the transverse mass
$m_{T}$ in the  $W \to l \nu$ decay this variable is sometimes
called the stransverse mass.     
For the decay of  two massive identical  
invisible particles  $m_{T2}$ is defined  as
$$
   m_{T2}({\mu}_N) \equiv \min_{\not{p_T^1}+\not{p_T^2}={p_T^{miss}}} 
\{max [m_{T}^{1}(p_{v}^{1},\not{p_{T}^1}, {\mu}_N),
m_{T}^{2}(p_{v}^{2},\not{p_{T}^2},{\mu}_N)]\}.$$
Here  $p_{v}^{1(2)}$ are  the sum of the momenta 
of the visible decay products
of a parent particles, ${\mu}_N$ is the trial mass parameter, namely,      
the mass of the LSP and 
$m_T(p_v,p_i,m_i) = m_v^2 + m_i^2 + 2(E_{vT}E_{iT} - \vec p_{vT}\vec p_{iT}),$
$E_{i(v)T}=\sqrt{\vec p_{i(v)T}^2 + m_{i(v)}^2}$.
The minimization is taken over all possible missing energy 
$p_{T}^{miss}$ splittings.
\begin{table}[]
\large {
\begin{center}
{\color{black}        
\begin{tabular}{||c||c||c||c||c||c||c||}     \hline \hline
$ m_{T2}\,[GeV] $ & 200 & 300  & 350 & 400 & 450 & 500 \\ \hline 
$S/B$ & 0.027 & 0.10  & 0.19 & 0.35 & 0.54 & 0.74 \\ \hline \hline 
\end {tabular}
\caption{The signal-to-background ratio for the  $\tilde q \bar {\tilde q}$ 
production in  pp interactions at $E_{CM}=10$~TeV and 1~fb$^{-1}$
of integrated luminosity estimated at different selection 
cuts for $m_{T2}$.}
}
\end{center}
}
\end{table}

In Fig.~3 (left panel) we demonstrate the $m_{T2}$ distribution for
events with  the sum of the transverse momenta  of two hardest jets 
greater than 500 GeV.
For this calculations we assume the massless LSP.                 
\begin{figure}[h!]                            
{\vskip -3.0cm}
{\hskip  0.0cm}  {\epsfxsize  6.0 truein \epsfbox{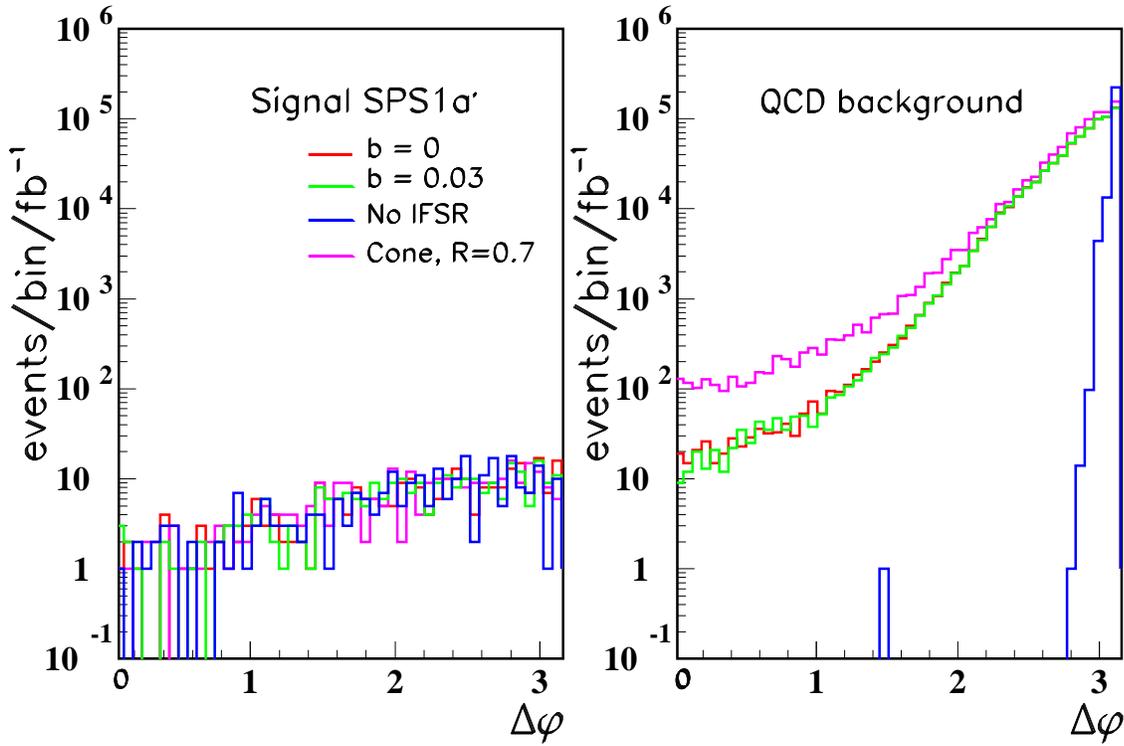}}
{\vskip -2.0cm}
\caption[]{Same as in Fig.~4 but for the difference of the 
azimuthal angles of two
hardest jets.}
\label{Norm}
\end{figure}

\begin{figure}[h!]                            
{\vskip -2.0cm}
{\hskip  0.0cm}  {\epsfxsize  6.0 truein \epsfbox{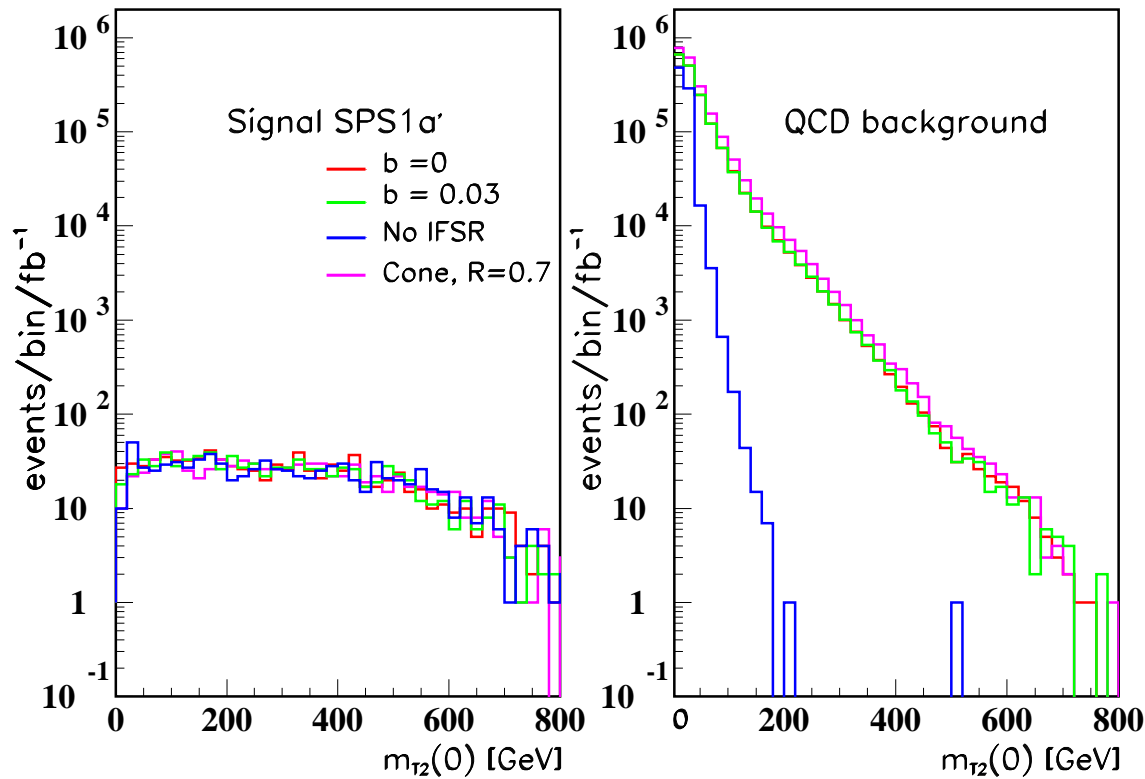}}
{\vskip -2.0cm}
\caption[]{Same as  in Fig.~4 but for $m_{T2}$ distribution.}                                  
\label{Norm}
\end{figure}

The S/B     for different  $m_{T2}$ cuts is presented 
in Table~2. The signal significance of 8.2 is
achievable using the $m_{T2}$ cut alone and $\sim$10 with
the combination of  $m_{T2}$ and $E_{T}^{miss}$. 
The tandem of $H_{T}^{miss}$ 
and $m_{T2}$ does  not lead  to improvement of the signal
significance.
\begin{figure}[h!]                            
{\vskip -3.5cm}
{\hskip  0.0cm}  {\epsfxsize  6.0 truein \epsfbox{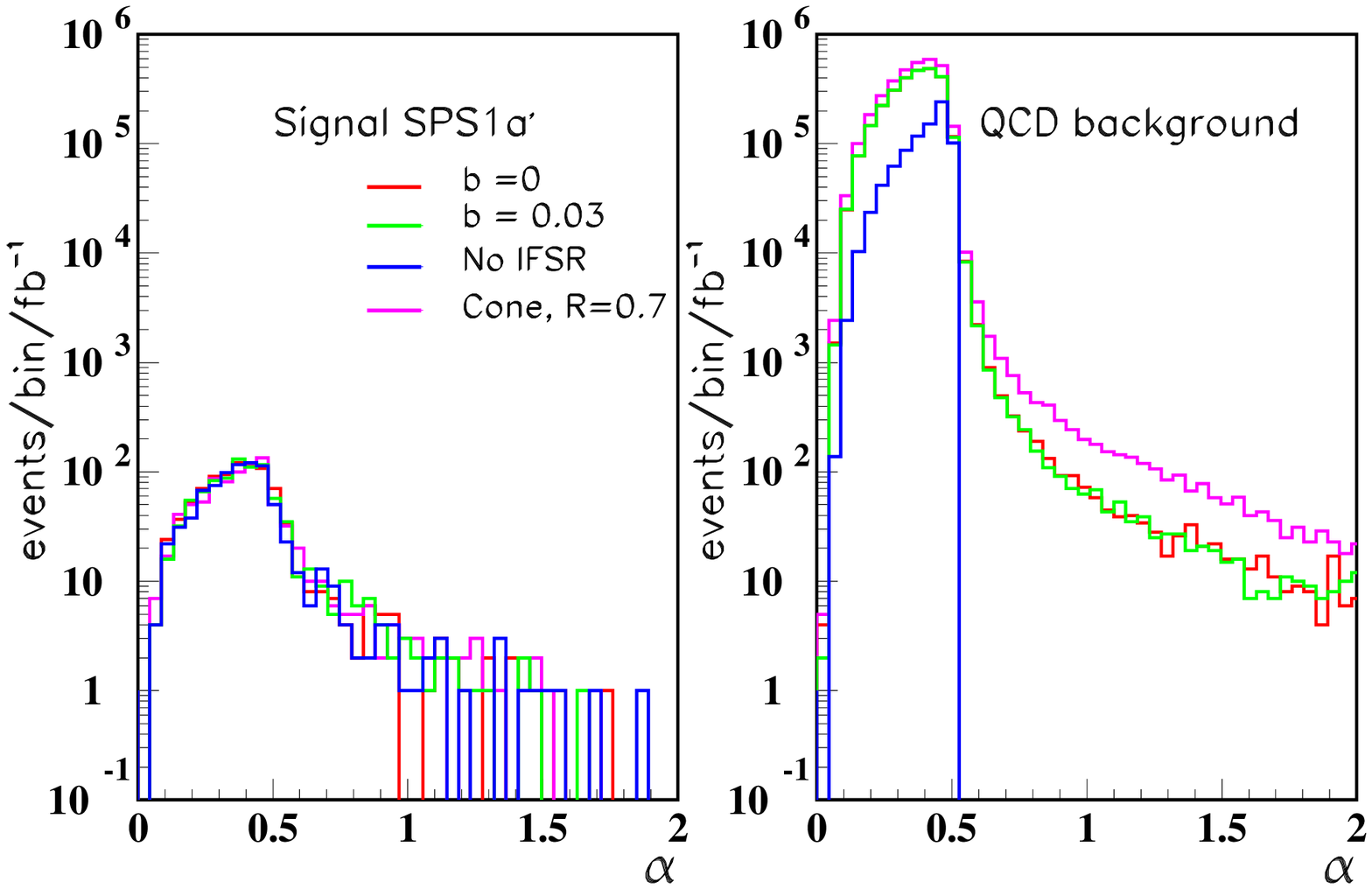}}
{\vskip -2.0cm}
\caption[]{Same as in Fig.~4 but for $\alpha$ distribution.}                   
\label{Norm}
\end{figure}

\begin{figure}[h!]
{\vskip -3.0cm}
{\hskip  0.0cm}  {\epsfxsize  6.0 truein \epsfbox{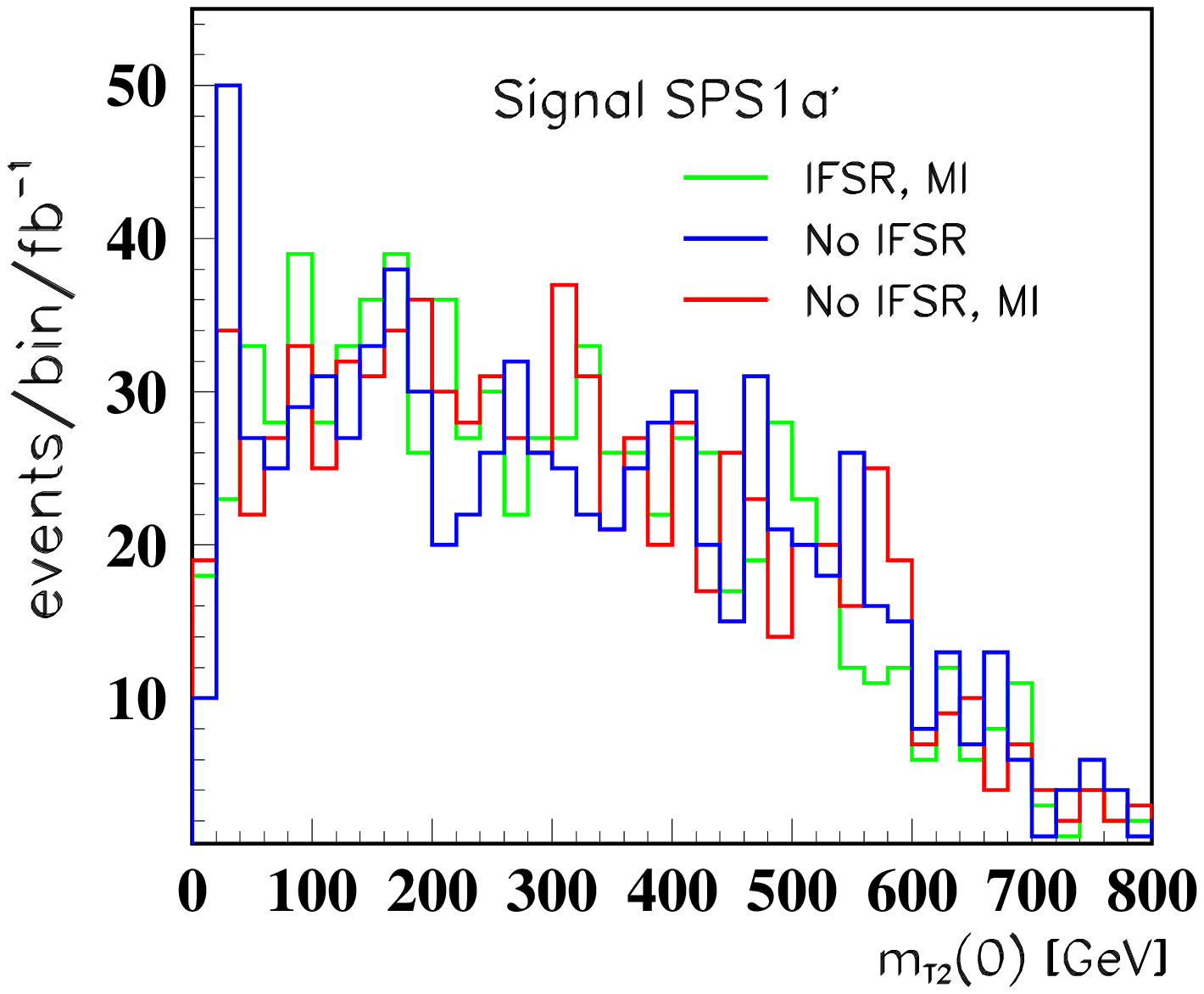}}
{\vskip -2.5cm}
\caption[]{Expected $m_{T2}$ distributions for signal events with initial- 
and final-state
radiation and multiple interactions  and without them   when 
the sum of the momenta of two hardest jets is greater than 500 GeV.
The data correspond to 1 fb$^{-1}$ of integrated
luminosity at $E_{CM}=10$ TeV.}
\label{Norm}
\end{figure}

We find that  0.25(0.4)fb$^{-1}$ of an integrated luminosity
when the LHC  runs    at  $E_{CM}$=10 TeV  
would provide observation of the $\tilde q \bar {\tilde q}$ signal
with 5 standard deviations of statistical uncertainty 
using the $E_{T}^{miss}(m_{T2})$ distribution.
Further, we estimate that a factor of $\sim$2.3 less integrated
luminosity is required at 14~TeV, and a factor of about 2.5 more 
at 7~TeV, to achieve the same signal sensitivity as at~10 TeV.

Recently, the new  measurable  $\alpha$\cite{rand} 
has been  proposed which can be used as a
signal/background discriminating variable  
in squark pair production. This  analysis does not explicitly
require the $E_{T}^{miss}$ cut.
The  variable $\alpha$  is defined as the ratio of the 
second hardest jet transverse momentum to  the invariant mass
$m_{inv}^{1,2}$ of two hardest jets
$$ \alpha = {p_{T}^2 \over m_{inv}^{1,2}}.$$

In Fig.~3 (right panel) we show the $\alpha$ distributions for 
the signal and background events. As one can see, the  clean
SUSY signal at  larger $\alpha$ values is not viable
due to a bigger tail of the 
QCD background distribution. The  $\alpha > 0.6$
signal is 45 times  weaker than the SM background.

To get further insight, in Figs.~4-7   we show 
what happens with the 
$E_{T}^{miss}$, $\Delta \phi$, $m_{T2}$ and $\alpha$  
distributions of the  signal and QCD  background events  when 
IFSR are  switched off in the simulated events. 
The impact of the constant 
term $b$ in the calorimeter energy resolution  formula 
 and jet definition
algorithm is also demonstrated in these figures. The cyan histogram
in the  figures corresponds to the cone jet definition
algorithm with  the cone radius $R=0.7$.
The signal event distributions are robust to 
all these influences. The presence of IFSR significantly 
increases the 
QCD background for  all these distributions                       
at   large values of the $E_{T}^{miss}$, $m_{T2}$ and $\alpha$ and
small values of $\Delta \phi$.
We can see that the missing transverse 
energy for the QCD background  at large values of $E_{T}^{miss}$ 
is less susceptible to the IFSR. The impact of the 
MI  on the $m_{T2}$ distribution of the
signal events is  demonstrated in Fig.~8.
One  can see that the upper edge of these  distributions          
indicates  to  the mass of the parent squark. The                  
experimental resolution leads to the smearing of the
correct mass point.

Finally, we remark that 
the behavior of the $\alpha$ distribution 
of the QCD background without IFSR in Fig.~7
looks surprisingly  like the ones obtained 
in Refs.~\cite{rand,oztu}.
\section{Conclusions}
$ $

We have demonstrated that squark pair production in 
pp collisions at the center-of-mass energy of
$E_{CM}=10$ TeV is a promising channel
for discovery of SUSY at the early stages of LHC running.
The our attention was focused on the distinctive characteristics
of  background subtractions in $\tilde q \bar {\tilde q}$ production
for the  mSUGRA benchmark point SPS1a' where the squarks are light.

With 1 fb$^{-1}$ of integrated luminosity,  strong evidence
for  a $\tilde q  \bar  {\tilde q}$ can appear 
with hard cuts on  the  missing
transverse energy and stransverse mass distributions.
In the mSUGRA model with a benchmark point SPS1a' it is
possible to reach  a signal significance ~ 5$\sigma$ 
of statistical uncertainty
with  0.4 fb$^{-1}$ of accumulated data when LHC 
operates at $E_{CM}=10$ TeV.

We find that initial- and final- state radiation forms
a crucial background for  the $\alpha$\cite{rand}  distribution
of the $\tilde q \bar {\tilde q}$ signal.
It should be hard   to pick out the
signal against  the background  in SUSY dijet events
using  the measured
$\alpha$  distribution even with a tandem
of other measurables.
The variable $\alpha$  not can provide an obvious
advantage for discovery of $\tilde q \bar{ \tilde q}$
as  was stated previously.

The careful validation of the whole richness of new physics  
at    the LHC will  be possible with improvement of our understanding
of the known physics.
}
      
}

\begin{thebibliography}{99}
\bibitem{susy1}
Yu.A. Golfand and E.P. Likhtman, JETP Lett. 13, 323 (1971) [Pisma Zh. Eksp. Teor.
 Fiz. 13, 452 (1971)];\\
D.V. Volkov and V.P. Akulov, JETP Lett. 16, 438 (1972) [Pisma Zh. Eksp. Teor. Fiz 
16, 621 (1972)];\\
J. Wess and B. Zumino, Phys. Lett. B49, 52 (1974).
\bibitem{ued1}
T. Appelquist, H.-C. Cheng and B.A. Dobrescu, Phys.Rev. D64, 035002 (2001);  \\
H.-C. Cheng, K.T. Matchev and M. Schmaltz, Phys.Rev. D66, 056006 (2002).
\bibitem{lith}
N. Arkani-Hamed, A.G. Cohen and H. Georgi, Phys.Lett. B513, 232 (2001).
\bibitem{faye}
P. Fayet, Phys.Lett. B69, 489 (1977);\\
G.R. Farrar and P. Fayet, Phys.Lett. B76, 575 (1978).
\bibitem{msug1}
A. Chamseddine, R. Arnowitt and P. Nath, Phys.Rev.Lett. 49, 970 (1982);\\
R. Barbieri, S. Ferrara and C. Savoy, Phys. Lett B119, 343 (1982); \\
L. Hall. J. Lykken and S. Weinberg, Phys.Rev. D27, 2359 (1983); \\
L. Alvarez-Gaume, J. Polchinski and M.B. Wise, Nucl.Phys. B221, 495 (1983).
\bibitem{bae2}
H. Baer, AIP Conf.Proc. 1200, 45 (2010), arXiv:0909.1515 [hep-ph].
\bibitem{cdfd0s}
CDF Collab., T. Aaltonen et al., Phys.Rev.Lett. 102, 121801 (2009);\\
D0 Collab., V.M. Abazov et al., Phys.Lett. B659, 856 (2008); 
Phys. Lett. B660, 449 (2008).
\bibitem{lep1}
ALEPH Collab., A. Heister et al., Phys.Lett. B537, 5 (2002); \\
DELPHI Collab., J. Abdallah et al., Eur.Phys.J. C31 421 (2003); \\
L3 Collab., P. Achard et al., Phys.Lett. B580, 37 (2004); \\
OPAL Collab., G. Abbiendi et al.,Phys.Lett. B545, 272 (2002);\\
http://lepsusy.web.cern.ch/lepsusy.                                            
\bibitem{hera}
G. Brandt, Recent HERA Results Sensitive to SUSY, arXiv:0809.3509 [hep-ex].
\bibitem{atl1}
ATLAS Collab., ATLAS Detector and Physics Performance Technical Design Report,
CERN-LHCC-99-14/15 (1999); JINST 3, S08003 (2008).
\bibitem{cms1}
CMS Collab.,  CERN-LHCC-2006-021; CMS-TDR-008-2; J.Phys. G34, 995 (2007).
\bibitem{lest}
C.G. Lester and D.J. Summers, Phys.Lett. B463, 99 (1999); A. Barr, C. Lester
and P. Stephens, J. Phys. G29, 2343 (2003); C. Lester and A. Barr, JHEP 0712,
102 (2007); A.J. Barr, B. Gripaios and C.G. Lester. JHEP 0802, 014 (2008).
\bibitem{rand}
L. Randall and D. Tucker-Smith, Phys.Rev.Lett. 101, 221803 (2008).
\bibitem{pyt1}
T. Sjostrand, S. Mrenna, and P. Skands, JHEP 0605, 026 (2006).
\bibitem{alpg}
M.L. Mangano, M. Moretti, F. Piccinini, R. Pittau, and A.D. Polosa, \\
JHEP 0307, 001 (2003).
\bibitem{sps1}
B.C. Allanach et al., Eur.Phys.J. C25, 113 (2002).
\bibitem{sphe}
W. Porod, Comput.Phys.Commun. 153, 275 (2003).
\bibitem{pump}
J. Pumplin et al., JHEP 0207, 012 (2002).
\bibitem{conw}
J. Conway, PGS-4 Simulation package for generic collider detectors, \\
http://www.physics.ucdavis.edu/${\sim}$conway/research/software/pgs/pgs.html
\bibitem{camp}
J. Campbell and R.K. Ellis, Phys.Rev. D65 113007 (2002); \\
J. Campbell, R.K. Ellis and D.L. Rainwater, Phys.Rev. D68 094021 (2003).
\bibitem{ditt}
S. Dittmaier, P. Uwer, S. Weinzierl, NLO QCD corrections to $pp \to t \bar t + jet+X$,
PoS RADCOR2007, 011 (2007), arXiv:0804.4389 [hep-ph].
\bibitem{wore}
M. Worek, Next-to-leading order $t \bar t$ plus jets physics with
HELAC-NLO, arXiv:1005.0306 [hep-ph].
\bibitem{prosp}
W. Beenakker, R. H\"opker, M. Spira, PROSPINO: A Program for the production
of supersymmetric particles in next-to-leading order QCD, 
hep-ph/9611232.
\bibitem{oztu}
T. Rommerskirchen, Searches for new phenomena at CMS and ATLAS,
arXiv:0905.4154 [hep-ex]; \\
N. Ozturk, Search for Supersymmetry Signatures at the LHC, arXiv:0910.2964 [hep-ex];  \\
A.D. Tapper, Early SUSY searches at the LHC, XXth Hadron Collider Physics Symposium,
November 16-20, 2009.\\
M. Pioppi, Search for SUSY at LHC in the first year of data-taking,
arXiv:0912.1189 [hep-ex]; \\
M.-H. Genest, Prospects for R-Parity Conserving SUSY searches at the LHC,
arXiv:0912.4378 [hep-ex].
\end{thebibliography}
\end{document}